%% file: main.tex
\documentclass[sigconf]{acmart}
\AtBeginDocument{%
  }

\setcopyright{acmlicensed}
\copyrightyear{2026}
\acmYear{2026}
\setcopyright{cc}
\setcctype{by}
\acmConference[CHI EA '26]{Extended Abstracts of the 2026 CHI Conference on Human Factors in Computing Systems}{April 13--17, 2026}{Barcelona, Spain}
\acmBooktitle{Extended Abstracts of the 2026 CHI Conference on Human Factors in Computing Systems (CHI EA '26), April 13--17, 2026, Barcelona, Spain}
\acmDOI{10.1145/3772363.3798907}
\acmISBN{979-8-4007-2281-3/2026/04}

\usepackage{booktabs}
\usepackage{multirow} 
\usepackage{subcaption}
\usepackage{float}

\begin{document}

\title{Seeing Your Mindless Face: How Viewing One’s Live Self Interrupts Mindless Short-Form Video Scrolling }


\author{Kyungjin Kim}
\affiliation{%
 \department{Department of Human-Artificial Intelligence Interaction}
 \institution{Sungkyunkwan University}
   \city{Seoul}
  \country{South Korea}
  }
\email{kjskkuis@g.skku.edu}
\orcid{0009-0008-0275-6318}

\author{Minjeong Kim}
\affiliation{%
 \department{Department of Human-Artificial Intelligence Interaction}
 \institution{Sungkyunkwan University}
  \city{Seoul}
  \country{South Korea}
  }
\email{minjeong5213@g.skku.edu}
\orcid{0009-0006-8176-1275}

\author{Soobeen Jeong}
\affiliation{%
  \department{Department of Human-Artificial Intelligence Interaction}
 \institution{Sungkyunkwan University}
  \city{Seoul}
  \country{South Korea}
}
\email{sb9916@g.skku.edu}
\orcid{0009-0001-8978-546X}

\author{Jiyeon So}
\affiliation{%
  \department{Department of Communication}
  \institution{Yonsei University}
  \city{Seoul}
  \country{South Korea}
  }
  \email{jso@yonsei.ac.kr}
  \orcid{0000-0003-2270-4490}

\author{Hayeon Song}
\authornote{Corresponding author}
\affiliation{%
  \department{Department of Human-Artificial Intelligence Interaction}
  \institution{Sungkyunkwan University}
  \city{Seoul}
  \country{Republic of Korea}}
\email{songhy@skku.edu}
\orcid{0000-0001-5951-8507}

\renewcommand{\shortauthors}{Kim et al.}

\begin{abstract}
The widespread, addictive consumption of short-form videos, which allegedly causes “brain rot,” has become an urgent public concern. This study proposes that self-related cues serve as an intrinsic, self-reflective strategy that enhances self-control over media overuse. We developed an app that de-immerses users by periodically displaying different self-related cues (live camera, selfie, name in text, and black screen) and tested their effects in a laboratory experiment (\textit{N}=84). Overall, findings show that self-related cues effectively disrupt mindless viewing, enabling users to voluntarily stop short-form video consumption. Interestingly, the black screen, intended as a control, elicited the greatest intention to use the app: Participants noted in the follow-up interview that they preferred the subtler reflection on a black screen over the explicit image from a live camera. The findings offer practical design guidelines for implementing self-awareness interventions in mobile contexts, including which modalities work best and how real-time contextual anchoring enhances effectiveness.
\end{abstract}


\begin{CCSXML}
<ccs2012>
   <concept>
       <concept_id>10003120.10003121.10011748</concept_id>
       <concept_desc>Human-centered computing~Empirical studies in HCI</concept_desc>
       <concept_significance>300</concept_significance>
       </concept>
 </ccs2012>
\end{CCSXML}

\ccsdesc[300]{Human-centered computing~Empirical studies in HCI}

\keywords{Short-form video consumption, Self-related cues, Objective self-awareness, Design friction, Digital wellbeing, De-immersion}
%% information and the body of the documentz, and typically spans the


\maketitle

\section{Introduction}
\input{1_Introduction}

\section{Related Work}
\input{2_RelatedWork}

\section{System Design}
\input{3_SystemDesign}

\section{Method}
\input{4_Method}

\section{Result}
\input{5_Result}

\section{Discussion}
\input{6_Discussion}

\section{Conclusion}
\input{7_Conclusion}

\begin{acks}
    This research was supported by the MSIT(Ministry of Science, ICT), Korea, under the Global Scholars Invitation Program(RS-2024-00459638) supervised by the IITP(Institute for Information \& Communications Technology Planning \& Evaluation) and the "Regional Innovation System \& Education (RISE)" through the Seoul RISE Center, funded by the Ministry of Education (MOE) and the Seoul Metropolitan Government. (2025-RISE-01-018-05)
\end{acks}

\bibliographystyle{ACM-Reference-Format}
\bibliography{reference}

\clearpage
\appendix
\input{Appendix}

\end{document}

%% file: 1_Introduction.tex
Short-form video content, brief clips typically lasting 20-40 seconds\cite{DemandSage2025}, has become central to global media consumption\cite{ZHANG2019101243}, with YouTube Shorts alone generating over 200 billion daily views\cite{youtube2025cannes}. This popularity, however, has come with a price: it has raised public concerns over “brain rot,” the Oxford Word of the Year 2024, referring to cognitive deterioration from overconsumption of trivial online content\cite{oxford2024brainrot}. The rapid context switching inherent in short-form videos significantly degrades prospective memory, impairing users' ability to remember and execute their intended actions\cite{10.1145/3544548.3580778}, including intentions to stop viewing or switch to other tasks.

To address this problem, various digital wellbeing tools have been proposed. However, many existing approaches rely on coercive strategies, such as app blocking\cite{10.1145/3314403, 10.1145/2858036.2858403}, which trigger psychological reactance\cite{10.1145/3313831.3376672, 10.1145/3613904.3642790}.  A more viable, effective alternative that is not repelling is needed to curb global overconsumption of short-form videos. To this end, we propose a shift from external coercion to internal self-regulation. Drawing on Objective self-awareness theory, which posits that directing attention inward motivates individuals to align their behavior with internal standards, we utilize self-related cues as a form of “design friction”. Prior Objective self-awareness research has employed diverse stimuli—ranging from physical mirrors to photographs and video feeds\cite{article1979, FEDEROFF1976336, ICKES1973202}—often treating these modalities as functionally interchangeable for inducing self-awareness. To identify  the effective intervention for mobile contexts, we systematically compared distinct modalities including visual cues (live camera, static selfie) and a text-based cue (user name). Additionally, a plain black screen was included as a control condition to evaluate the distinct impact of self-representation.

Based on our findings, this study makes two key contributions. First, we demonstrate that self-related cues effectively disrupt unconscious viewing patterns, enabling users to voluntarily stop short-form video consumption—not through external blocking or restriction, but through heightened self-awareness that prompts autonomous disengagement. Second, we provide specific design guidelines for implementing self-awareness interventions in mobile contexts, identifying which modalities are most effective, when explicit exposure becomes counterproductive, and how real-time contextual anchoring enhances effectiveness.

%% file: 2_RelatedWork.tex
\subsection{Design Friction for Digital Wellbeing}
\citet{10.1145/2851581.2892410} introduced “microboundaries”—subtle design frictions that disrupt mindless automaticity without strictly forbidding usage. Recent studies have validated that non-coercive interventions, ranging from brief temporal frictions combined with dismissal options\cite{doi:10.1073/pnas.2213114120} to LLM-generated context-aware persuasive messages\cite{10.1145/3613904.3642790}, effectively support user agency and self-regulation\cite{10.1145/3613904.3642370}. While these approaches restore the “self” to the driver’s seat, the efficacy of using self-related cues (e.g., visual self-image) in doing so remains largely unexplored. We address this gap by examining how “seeing oneself” functions as a mechanism to interrupt immersion and regain voluntary control.

\subsection{Objective self-awareness}
Research confirms the brain’s heightened sensitivity to self-related cues\cite{CARMODY2006153,https://doi.org/10.1002/hbm.22703,QIN20111221}. According to Objective self-awareness theory, such stimuli direct attention inward, promoting self-evaluation against internal standards and motivating corrective action\cite{duval1972theory,Silvia2001Objective}. However, prior work has largely relied on physical mirrors\cite{GOVERN2001366} or pre-recorded videos\cite{PICCO2025421}, leaving the effects of real-time digital self-representations in mobile contexts unexplored. This study addresses this gap by systematically comparing diverse self-related cues as forms of design friction within a controlled mobile short-form video context.

%% file: 3_SystemDesign.tex
\textit{SelfStop} is an Android application designed to interrupt unconscious short-form video consumption by periodically displaying self-related cues during YouTube Shorts viewing. The system replicates the standard YouTube Shorts interface—featuring a full-screen vertical video player with swipe-based navigation and voluntary session termination—to preserve ecological validity. Participants accessed our app and logged in using the same Google account they typically use for YouTube, allowing them to view a personalized recommendation feed based on their prior watch history.

Based on pilot testing ($N$=5), the system triggers a 5-second full-screen intervention after every 20th video, balancing viewing continuity with periodic cognitive interruption. In the absence of established prior work specifying appropriate intervention intervals and durations, these design choices were determined through pilot testing to ensure consistency across participants and enable controlled comparison of intervention effects. Upon triggering, the video stream is interrupted by a full-screen overlay displaying one of four experimental conditions: (1) Black Screen (control), (2) Live Camera (real-time self-image via front camera), (3) Selfie (pre-captured static photo), or (4) Name in Text (participant's name displayed as text). After 5 seconds, normal viewing resumes automatically (Figure~\ref{fig:process_flow}). See Appendix~\ref{app:system} for system design and privacy details.

%%figure
\begin{figure}[h]
  \centering
  \includegraphics[width=\linewidth]{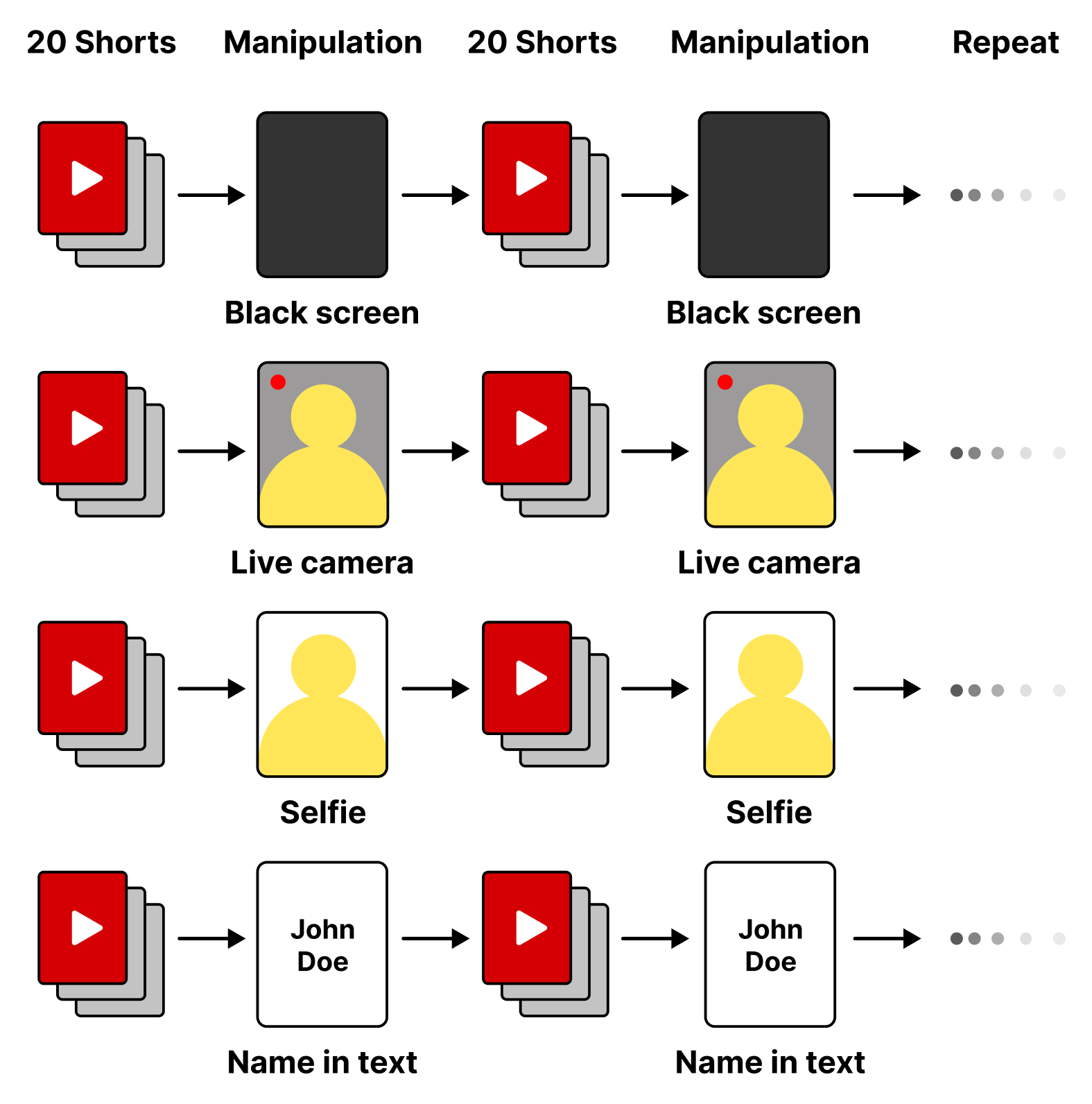} 
  \caption{Participants encountered a five-second intervention after every 20 short-form videos. The intervention involved one of four types of self-related cues: black screen, live camera, selfie, or name in text.}
    \label{fig:process_flow}

   \Description{This flowchart depicts how interventions were administered during the experiment. After every 20 Shorts, participants were exposed for five seconds to one of four self-related cues—black screen, live camera, selfie, or name in text—according to their randomly assigned condition. After each intervention, the application returned to Shorts viewing, and this cycle was repeated throughout the session.}
\end{figure}

%% file: 4_Method.tex
\subparagraph{\textbf{\textit{Participants.}}} To test the relative effects of different self-related cues, we conducted a lab experiment with 84 participants (21 per condition; 71\% female, $M_{age}$ = 24.32, $SD = 3.28$). On an average day, participants viewed YouTube for 124.01 minutes ($SD=91.25$),  as recorded by their phones' logging systems. According to self-report, 33.6\% of this time was spent watching Shorts.

\subparagraph{\textbf{\textit{Procedure.}}} Participants were invited to a lab and instructed to watch Shorts as they would in their everyday lives with a smartphone, on which \textit{SelfStop}, the app developed for this study, was installed. They were told to watch Shorts as much as they want within a 30-minute limit. To ensure privacy and simulate a natural setting, partitions, cushions, and blankets were provided. After the viewing session, they were asked to fill out an online questionnaire and participate in an interview. See Figure 4 in Appendix~\ref{sec:setting} for lab setting details.

\subparagraph{\textbf{\textit{Measurement.}}} We collected behavioral logs, self-reported questionnaire data, and qualitative interview data. Behavioral measures included the number of Shorts watched. Self-reported measures (7-point Likert scales) assessed Objective self-awareness (9 items, $\alpha=.81$)\cite{GOVERN2001366}, Attitude (3 items, $\alpha=.89$)\cite{PARK2024102594}, Satisfaction (2 items, $\alpha=.91$)\cite{662ada62-a334-3561-89a0-428c9f338300}, Perceived Usefulness (1 item)\cite{WITTE01101996}, Self-efficacy (4 items, $\alpha=.80$)\cite{Ralf2009}, and Intention to Use (2 items, $\alpha=.94$)\cite{doi:10.1287/mnsc.46.2.186.11926}. See Appendix~\ref{appc:Survey Items} for a complete list of survey items. Qualitative data were gathered through semi-structured interviews with 20 randomly selected participants (5 per condition). The interview protocol explored two main areas: (1) participants' experiences using the app, and (2) prior experience with other similar apps.

\subparagraph{\textbf{\textit{Data analysis.}}} Behavioral and survey data were analyzed using one-way ANCOVA with baseline daily YouTube viewing time as a covariate. For qualitative data, all interviews were transcribed and analyzed using a bottom-up inductive thematic analysis approach. Three researchers independently conducted open coding, then held calibration meetings to consolidate patterns into themes and resolve discrepancies through discussion.

%% file: 5_Result.tex
\subsection{Quantitative Results}
Although technical validation confirmed the system’s stability and functionality, behavioral log data were missing for 10 participants due to isolated logging synchronization errors. Thus, the behavioral analyses were conducted with the data from 74 participants. Since daily viewing time significantly differed across conditions, $F(3,80)=2.89$, $p=.04$, it was treated as a covariate in subsequent analyses.

Prior to conducting the ANCOVA, fundamental statistical assumptions were verified. Levene’s test confirmed the homogeneity of variances for most variables; while the assumption was not strictly met for Intention to Use, the analysis proceeded as ANCOVA is generally robust to such violations when group sizes are balanced. Normality was confirmed via visual inspection of histograms, which indicated that the distributions for all key variables were reasonably normal. Furthermore, the interaction between the independent variable and the covariate  was not statistically significant ($p = .39$), confirming that the assumption of homogeneity of regression slopes was satisfied.

\begin{figure}[ht]
    \centering
    \includegraphics[width=\linewidth]{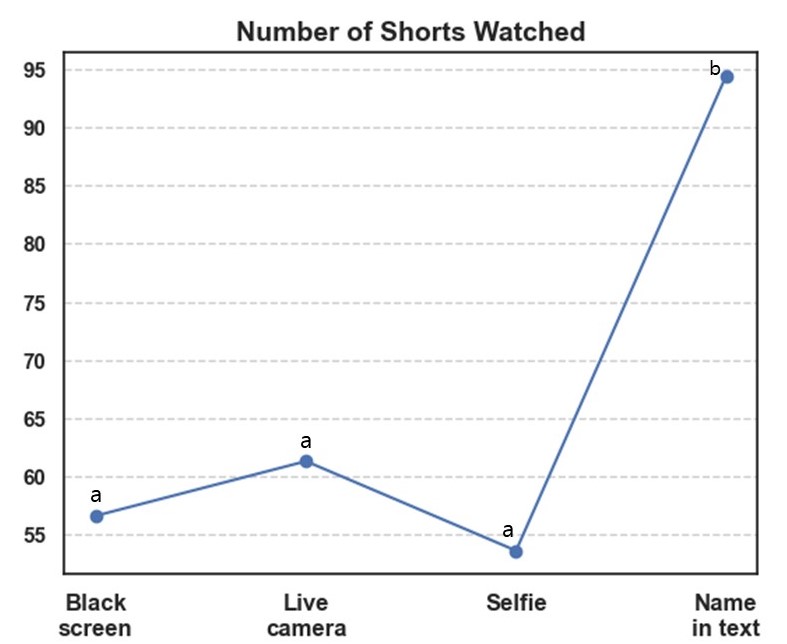}
    \caption{Mean number of Shorts watched across the four intervention conditions. Means with different letters differ significantly at $p < .05$.}
    \Description{This line graph compares the number of Shorts watched across four conditions: black screen, live camera, selfie, and name in text. The Y-axis represents the number of Shorts watched (40–100). The selfie condition had the lowest mean (about 54), followed by black screen (about 57) and live camera (about 61), whereas the name in text condition showed the highest mean (about 92). Name in text was significantly higher than black screen and selfie ($p < .05$).}
    \label{fig:shorts}
\end{figure}

 An ANCOVA revealed a significant main effect of the condition on the number of Shorts watched, $F(3, 69) = 6.07, p < .001, \eta_p^2 = .21$. Post-hoc comparisons with Bonferroni correction indicated that participants in the name in text condition ($M = 94.44, SD = 44.62$) viewed significantly more Shorts than those in the black screen ($M = 56.61, SD = 26.53, p = .006$), live camera ($M = 61.30, SD = 24.64, p = .02$), and selfie ($M = 53.61, SD = 28.21, p = .002$) conditions. There were no significant differences in the number of Shorts watched between the black screen, live camera, and selfie conditions. See Figure~\ref{fig:shorts}. 

\begin{figure}[ht]
    \includegraphics[width=\linewidth]{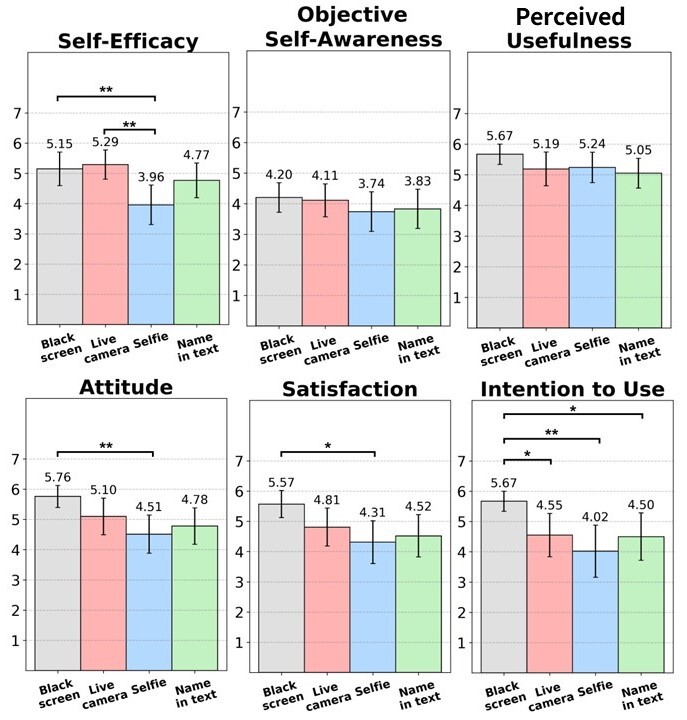}
    \caption{Mean self-reported ratings of user experiences across the four intervention conditions. Asterisks indicate significant differences (* $p < .05$, ** $p < .01$).}
    \Description{This bar graph shows self-reported outcomes across four intervention conditions. The Y-axis represents mean values on a seven-point Likert scale. Self-efficacy is highest in live camera and lowest in selfie. Objective self-awareness shows no significant differences. Attitude, Satisfaction, and Intention to Use are higher in black screen and lower in selfie. Usefulness remains stable across conditions.}
    \label{fig:survey}
\end{figure}

Mixed results were observed across different self-reported measures. Attitude toward the application differed significantly across conditions, $F(3, 79) = 3.443, p = .021,  \eta^2=.12$. Participants in the black screen condition ($M = 5.76, SD = 0.80$) reported significantly more positive attitudes than those in the selfie condition ($M = 4.51, SD = 1.39, p = .009$). Similarly, Satisfaction showed a significant effect, $F(3, 79) = 3.002, p = .035, \eta^2 =.10$, with black screen ($M = 5.57, SD = 0.98$) yielding higher Satisfaction than selfie ($M = 4.31, SD = 1.56, p = .03$). Self-efficacy also varied significantly by condition, $F(3, 79) = 4.398, p = .006, \eta^2 =.14$. Both live camera ($M = 5.37, SD = 1.06, p = .002$) and black screen ($M = 5.17, SD = 1.22, p = .008$) conditions enhanced Self-efficacy more than the selfie condition ($M = 3.89, SD = 1.43$).

Additionally, Intention to Use differed across conditions, $F(3,79)=3.96, p=.01, \eta^2=.13$, with the black screen condition ($M = 5.67, SD = 0.73$) yielding higher Intention than live camera ($M = 4.55, SD = 1.57, p = .03$), name in text ($M = 4.50, SD = 1.72, p = .04$), and selfie ($M = 4.02, SD = 1.89, p = .005$).
Finally, no significant main effect was found for Objective self-awareness, $F(3,79) = 0.328, p = .805, \eta^2 = .012$ and Perceived Usefulness, $F(3, 79) = 1.297, p = .281, \eta_p^2 = .047$. See Figure~\ref{fig:survey}.

\subsection{Qualitative Results}
\subsubsection{From Immersion to Objective self-awareness: The “\textit{Hyunta}” Experience}
Across all conditions, participants experienced Objective self-awareness upon exposure to self-related cues. This led them to recognize discrepancies between their actual and expected selves—and critically, to voluntarily stop viewing rather than being forcibly blocked. L03\footnote{Each participant was assigned an ID composed of a letter and a number. The letter indicated the intervention condition to which the participant was assigned: L represented the live camera condition, S the selfie condition, T the name in text condition, and B the black screen condition.} stated: “\textit{I saw myself with a vacant gaze, just expressionless, not actually enjoying it},” realizing that although they believed they were watching Shorts for enjoyment, their facial expression suggested otherwise.
Notably, this self-awareness manifested in a distinct and intense form that participants referred to as “\textit{Hyunta}”—a Korean term meaning a sudden “reality hit,” in which one abruptly snaps out of deep immersion and becomes acutely aware of their current situation. Unlike Objective self-awareness as a gradual reflective process, “\textit{Hyunta}” captures an immediate and jarring break from immersion. S25 reported, “\textit{The sudden reality hit [“Hyunta”] felt like coming back to reality after being in a virtual world.}”

Beyond attention redirection, \textit{Hyunta} motivated participants to stop viewing voluntarily. "\textit{Seeing my face brought on a hyunta... made me think I should stop watching}" (L30). "\textit{It interrupted the endless flow of Shorts, breaking my immersion and making me aware of reality, which made me want to stop watching}" (B21). As T24 noted, the system \textit{"guides you to regulate yourself through awareness, rather than controlling you with an app lock format"}—self-initiated disengagement, not coercive blocking.
However, explicit self-exposure via live camera or a selfie, while effective at triggering \textit{Hyunta}, often elicited surprise, discomfort, and embarrassment (L04). “\textit{Showing a selfie without warning provoked shame and felt unpleasant}” (S30). 

\subsubsection{The “Black Mirror” Effect: An Unexpected Finding}
Unexpectedly, the black screen condition—originally designed as a control—elicited strong self-awareness effects. The majority of participants reported seeing their faces subtly reflected on the black screen, thereby inadvertently experiencing Objective self-awareness. “\textit{When the black screen pops up, I see my reflection and think, ‘How much time has passed?’ Then I realized I was watching a one-minute video, but 20 or 30 minutes have gone by}” (B35).
Participants reported that the black screen enabled real-time self-reflection without the explicit self-exposure of a live camera or a selfie. B02 stated: “\textit{I could see just my face on the black screen} [...] \textit{I was watching something funny, but then I get “hyunta” and thought, ‘Okay, I should stop now.’}” The same participant added: “\textit{If it had shown something else, it might have been annoying. But since it just shows a black screen, it really hit me} [...] \textit{it's something I could easily incorporate into daily life},” positively evaluating both effectiveness and acceptability.
This subtle reflection induced lower psychological reactance and higher acceptance compared to explicit self-image interventions. 

\subsubsection{Static Cues: Diverging Outcomes}
Qualitative findings suggest that both static cues failed to connect with the user’s current context. Participants noted a lack of contextual relevance and said “\textit{I'm not sure if just showing a name is sufficient}”(T03). “\textit{It felt like a YouTube ad}”(T25).
Selfie similarly failed to convey contextual connection due to a lack of real-time reflection. S03 and S22 noted that “\textit{it showed the initial photo rather than capturing the current moment.” “Only the selfie appeared with nothing else, so it had no [effect] at all}”(S23).
This limitation of static cues contrasts sharply with the advantages of real-time reflection from a black screen and a live camera. Seeing one’s present mindless face enabled participants to clearly recognize their current situation, achieving both high effectiveness and acceptability.

%% file: 6_Discussion.tex
This study demonstrates that self-related cues used as design friction can effectively reduce short-form video overconsumption by triggering Objective self-awareness without coercive control. These results suggest a new paradigm for digital wellbeing—one that empowers users through self-awareness rather than restricting them through external force. 

The contributions of this study are as follows: First, this study proposes a paradigm shift from external coercion to internal self-regulation by redefining the smartphone screen: not as a “window” into content, but as a “mirror” for self-reflection. This transformation triggers “\textit{Hyunta}” (Reality Hit). Theoretically, this extends Objective self-awareness theory to mobile contexts. Crucially, this mirror-like friction shifts the locus of control: unlike coercive digital wellbeing interventions \cite{10.1145/2858036.2858403, 10.1145/3314403}, our 5-second self-related cues disappeared automatically, and cessation occurred only when participants voluntarily chose to stop. Thus, this work lays the foundation for interventions that foster autonomous, conscious regulation over passive consumption.

Second, interestingly, black screen yielded the highest Intention to Use, Attitude, and Satisfaction across all conditions—a finding that warrants theoretical attention. Although initially intended as a control, the black screen functioned as an active reflective surface: users perceived their own dim reflections implicitly, without the overt self-confrontation that a live camera imposed. Prior work suggests that explicit self-representations can elicit embarrassment and psychological reactance\cite{morita2014anterior,kim2024voice}, which likely explains why live camera underperformed despite its stronger self-salience. The black screen, by contrast, delivered self-related cues subtly—bypassing resistance to persuasive intent through implicit persuasion. This aligns with recent social media research showing that implicit persuasion outperforms explicit approaches, as the latter can feel controlling and burdensome\cite{LIAO2024,Bhattacharjee2025} Notably, the fact  that even this ostensibly passive condition produced meaningful behavioral effects suggests that effective design friction lies not in the intensity of self-exposure, but in the equilibrium between explicitness and ambiguity.

Third, our findings highlight that self-representation must be grounded in real-time relevance. Although the static selfie condition resulted in the fewest Shorts viewed, it scored lowest on key user experience metrics, including Self-efficacy and Intention to Use. Qualitative feedback elucidates this paradox: participants felt dissonance because pre-captured images appeared “out of the blue” without real-time context. In contrast, the live camera and black screen maintained real-time relevance, creating a link to the user’s current state. Thus, for design friction to be effective, it requires contextual anchoring in the “Here and Now,” bridging the gap between the digital and physical self.

This paper has several limitations and plans for future research. Future research should extend ecological validity through longitudinal field deployments, moving beyond our 30-minute lab session with young adults. This is essential to examine long-term sustainability and determine whether repeated exposure leads to desensitization or effectively mitigates habituation. Additionally, we plan to optimize the intervention by systematically varying frequency and duration, overcoming the limitations of the fixed 20-video interval used in this study. Finally, future systems must be context-aware to ensure usability across diverse environments. For instance, low ambient light obscures physical reflections, rendering visual self-cues ineffective, whereas public usage requires adaptation, as explicit self-cues can precipitate social anxiety due to users’ heightened sensitivity to self-representation \cite {kim2024voice}. Therefore, next-generation designs should dynamically adjust intervention modalities based on ambient lighting and social context.

%% file: 7_Conclusion.tex
This study identifies an effective approach for interrupting unconscious short-form video consumption through the use of self-related cues as design friction. Among the tested modalities, the black screen emerged as a particularly promising intervention, inducing self-awareness with relatively low psychological resistance compared to more intrusive self-representation methods. By subtly reintroducing the self into the viewing experience, this approach can interrupt habitual consumption without relying on coercive restrictions. In addition, our findings suggest that, in the user experience of self-related cues, the liveness of one’s own appearance is an important factor in eliciting self-awareness. These findings suggest that carefully designed self-related cues may provide a viable pathway for supporting voluntary disengagement from short-form media.

%% file: Appendix.tex
\clearpage
\onecolumn
\section{Privacy and Technical Implementation}
\label{app:system}
\subparagraph{\textbf{\textit{Privacy and Ethical Safeguards.}}}
Given the sensitivity of camera-based interventions, \textit{SelfStop} was designed with a privacy-first architecture. Live camera feeds were rendered only locally without recording or transmission, and selfies were stored temporarily on-device and automatically deleted at session end. All behavioral data were anonymized using randomly generated user IDs, and personally identifiable information was not stored in the persistent database.
\subparagraph{\textbf{\textit{System Validation.}}}
Prior to the main experiment, the system underwent technical validation confirming logging accuracy and stable operation during 30-minute continuous streaming sessions. Minor usability adjustments were made to align swipe sensitivity and video loading behavior with user expectations for commercial short-form video platforms.

\section{Lab setting}
\label{sec:setting}
\begin{figure} [h!]
  \centering
  \includegraphics[width=0.9\linewidth]{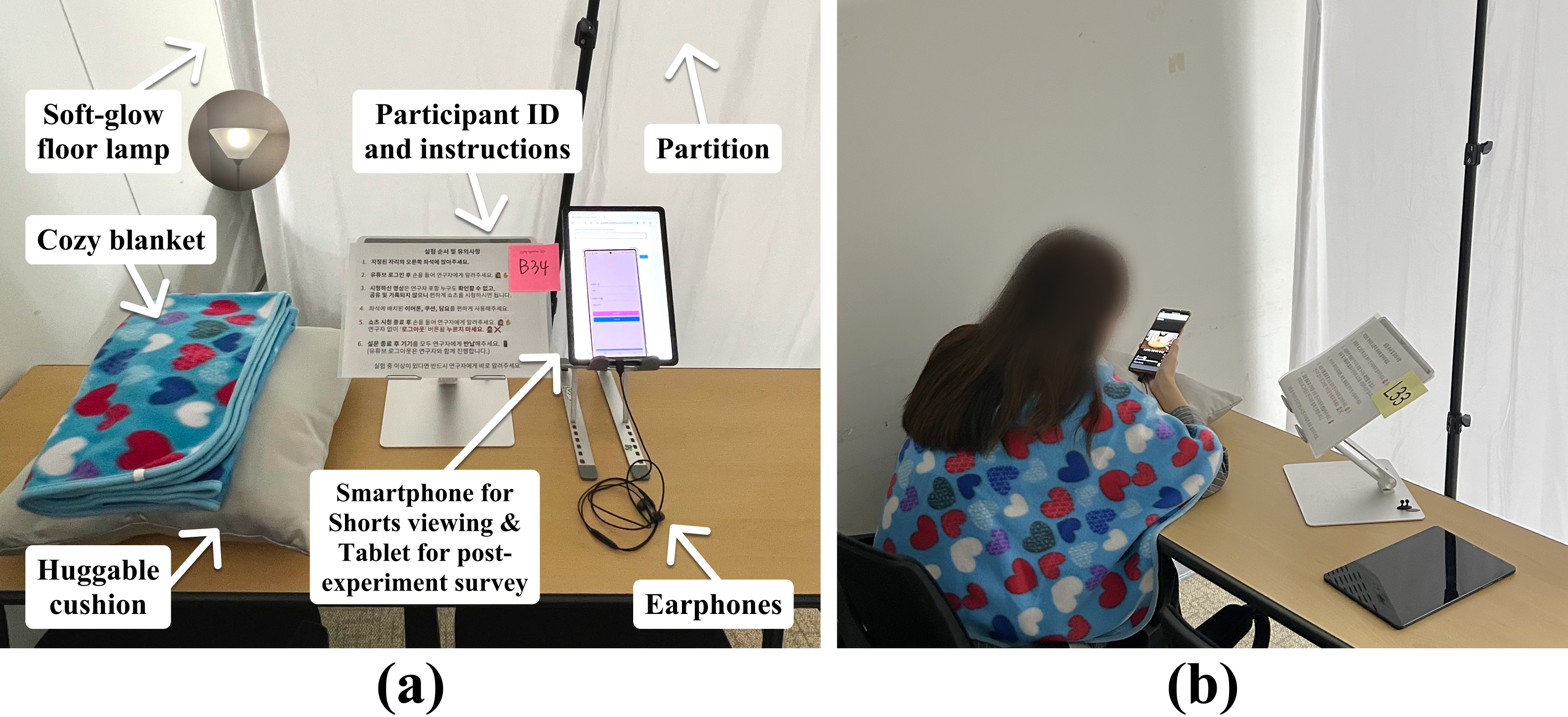}
  \caption{(a) The environment setup, including a partition, a soft-glow floor lamp, a cozy blanket, a huggable cushion, earphones, smartphone for Shorts viewing, and a tablet for the post-experiment survey.
(b) A participant participating in the experiment.}
  \Description{Panel (a) depicts the experimental environment: a partition, a soft-glow floor lamp, a cozy blanket, a huggable cushion, and earphones arranged on a desk. A participant ID sheet with instructions is placed on a stand, and two devices are provided: a smartphone for Shorts viewing and a tablet for completing the survey. Panel (b) shows a participant using the setup. The participant is seated at a desk, wrapped in a blanket, holding the smartphone to watch Shorts. A partition surrounds the desk to block the participant from others’ view, ensuring privacy during the session. The tablet and instruction sheet are also visible on the desk.}
  \label{fig:labsetting}
\end{figure}

\clearpage

\section{Survey Items}
\label{appc:Survey Items}
\begin{table}[h!] 
\caption{Measurement Items Used in the Study}
\small
\begin{tabular}{p{3.5cm} p{12cm}}
\toprule
Measure & Item (7-point Likert) \\
\midrule
Objective self-awareness\cite{GOVERN2001366}
& When a notification is displayed, I am keenly aware of everything in my environment. \\

& When a notification is displayed, I am conscious of my inner feelings. \\

& When a notification is displayed, I am concerned about the way I present myself. \\

& When a notification is displayed, I am self-conscious about the way I look. \\

& When a notification is displayed, I am conscious of what is going on around me. \\

& When a notification is displayed, I am reflective about my life. \\

& When a notification is displayed, I am concerned about what other people think of me. \\

& When a notification is displayed, I am aware of my innermost thoughts. \\

& When a notification is displayed, I am conscious of all objects around me. \\

\addlinespace

Attitude\cite{PARK2024102594}
& \textit{SelfStop} is a beneficial tool for me (ineffective \dots\ effective). \\

& \textit{SelfStop} helps me effectively achieve my purpose (useless \dots\ useful). \\

& \textit{SelfStop} helps me effectively achieve my purpose (bad \dots\ good). \\

\addlinespace

Satisfaction\cite{662ada62-a334-3561-89a0-428c9f338300}
& All things considered, I am very satisfied with \textit{SelfStop}. \\

& Overall, my interaction with \textit{SelfStop} is very satisfying. \\

\addlinespace

Perceived Usefulness\cite{WITTE01101996}
& Using \textit{SelfStop} is an effective way to reduce Shorts consumption. \\

\addlinespace

Self-efficacy\cite{Ralf2009}
& I can manage to reduce watching Shorts using \textit{SelfStop}, even when I have worries and problems. \\

& I can manage to reduce watching Shorts using \textit{SelfStop}, even if I feel depressed. \\

& I can manage to reduce watching Shorts using \textit{SelfStop}, even when I feel tense. \\

& I can manage to reduce watching Shorts using \textit{SelfStop}, even when I am bored. \\

\addlinespace

Intention to Use\cite{doi:10.1287/mnsc.46.2.186.11926}
& If possible, I intend to use \textit{SelfStop}. \\

& Given that I have access to \textit{SelfStop}, I predict that I would use it. \\

\bottomrule
\end{tabular}
\end{table}

%% file: main.bbl

\begin{thebibliography}{31}

%%% the Web address.

\ifx \showCODEN    \undefined \def \showCODEN     #1{\unskip}     \fi
\ifx \showISBNx    \undefined \def \showISBNx     #1{\unskip}     \fi
\ifx \showISBNxiii \undefined \def \showISBNxiii  #1{\unskip}     \fi
\ifx \showISSN     \undefined \def \showISSN      #1{\unskip}     \fi
\ifx \showLCCN     \undefined \def \showLCCN      #1{\unskip}     \fi
\ifx \shownote     \undefined \def \shownote      #1{#1}          \fi
\ifx \showarticletitle \undefined \def \showarticletitle #1{#1}   \fi
\ifx \showURL      \undefined \def \showURL       {\relax}        \fi
% invisible to TeX
\providecommand\bibfield[2]{#2}
\providecommand\bibinfo[2]{#2}
\providecommand\natexlab[1]{#1}
\providecommand\showeprint[2][]{arXiv:#2}

\bibitem[Bhattacharjee et~al\mbox{.}(2025)]%
        {Bhattacharjee2025}
\bibfield{author}{\bibinfo{person}{Debashree~Roy Bhattacharjee}, \bibinfo{person}{Debasis Pradhan}, \bibinfo{person}{Abhisek Kuanr}, {and} \bibinfo{person}{Neeru Malhotra}.} \bibinfo{year}{2025}\natexlab{}.
\newblock \showarticletitle{Perfectly Imperfect: How Body-Positive Advertisements in Social Media Foster Consumer Engagement}.
\newblock \bibinfo{journal}{\emph{Journal of Advertising}} \bibinfo{volume}{54}, \bibinfo{number}{1} (\bibinfo{year}{2025}), \bibinfo{pages}{79--98}.
\newblock
\href{https://doi.org/10.1080/00913367.2024.2306422}{doi:\nolinkurl{10.1080/00913367.2024.2306422}}


\bibitem[Carmody and Lewis(2006)]%
        {CARMODY2006153}
\bibfield{author}{\bibinfo{person}{Dennis~P. Carmody} {and} \bibinfo{person}{Michael Lewis}.} \bibinfo{year}{2006}\natexlab{}.
\newblock \showarticletitle{Brain activation when hearing one's own and others' names}.
\newblock \bibinfo{journal}{\emph{Brain Research}} \bibinfo{volume}{1116}, \bibinfo{number}{1} (\bibinfo{year}{2006}), \bibinfo{pages}{153--158}.
\newblock
\showISSN{0006-8993}
\href{https://doi.org/10.1016/j.brainres.2006.07.121}{doi:\nolinkurl{10.1016/j.brainres.2006.07.121}}


\bibitem[Chiossi et~al\mbox{.}(2023)]%
        {10.1145/3544548.3580778}
\bibfield{author}{\bibinfo{person}{Francesco Chiossi}, \bibinfo{person}{Luke Haliburton}, \bibinfo{person}{Changkun Ou}, \bibinfo{person}{Andreas~Martin Butz}, {and} \bibinfo{person}{Albrecht Schmidt}.} \bibinfo{year}{2023}\natexlab{}.
\newblock \showarticletitle{Short-Form Videos Degrade Our Capacity to Retain Intentions: Effect of Context Switching On Prospective Memory}. In \bibinfo{booktitle}{\emph{Proceedings of the 2023 CHI Conference on Human Factors in Computing Systems}} (Hamburg, Germany) \emph{(\bibinfo{series}{CHI '23})}. \bibinfo{publisher}{Association for Computing Machinery}, \bibinfo{address}{New York, NY, USA}, Article \bibinfo{articleno}{30}, \bibinfo{numpages}{15}~pages.
\newblock
\showISBNx{9781450394215}
\href{https://doi.org/10.1145/3544548.3580778}{doi:\nolinkurl{10.1145/3544548.3580778}}


\bibitem[Cox et~al\mbox{.}(2016)]%
        {10.1145/2851581.2892410}
\bibfield{author}{\bibinfo{person}{Anna~L. Cox}, \bibinfo{person}{Sandy~J.J. Gould}, \bibinfo{person}{Marta~E. Cecchinato}, \bibinfo{person}{Ioanna Iacovides}, {and} \bibinfo{person}{Ian Renfree}.} \bibinfo{year}{2016}\natexlab{}.
\newblock \showarticletitle{Design Frictions for Mindful Interactions: The Case for Microboundaries}. In \bibinfo{booktitle}{\emph{Proceedings of the 2016 CHI Conference Extended Abstracts on Human Factors in Computing Systems}} (San Jose, California, USA) \emph{(\bibinfo{series}{CHI EA '16})}. \bibinfo{publisher}{Association for Computing Machinery}, \bibinfo{address}{New York, NY, USA}, \bibinfo{pages}{1389–1397}.
\newblock
\showISBNx{9781450340823}
\href{https://doi.org/10.1145/2851581.2892410}{doi:\nolinkurl{10.1145/2851581.2892410}}


\bibitem[DemandSage(2025)]%
        {DemandSage2025}
\bibfield{author}{\bibinfo{person}{DemandSage}.} \bibinfo{year}{2025}\natexlab{}.
\newblock \bibinfo{booktitle}{\emph{YouTube Shorts Statistics (2025)}}.
\newblock
\newblock
\shownote{Accessed: 2025-09-09}.


\bibitem[Duval et~al\mbox{.}(1979)]%
        {article1979}
\bibfield{author}{\bibinfo{person}{Shelley Duval}, \bibinfo{person}{Virginia Duval}, {and} \bibinfo{person}{Robert Neely}.} \bibinfo{year}{1979}\natexlab{}.
\newblock \showarticletitle{Self-focus, felt responsibility, and helping behavior}.
\newblock \bibinfo{journal}{\emph{Journal of Personality and Social Psychology}}  \bibinfo{volume}{37} (\bibinfo{date}{10} \bibinfo{year}{1979}), \bibinfo{pages}{1769--1778}.
\newblock
\href{https://doi.org/10.1037/0022-3514.37.10.1769}{doi:\nolinkurl{10.1037/0022-3514.37.10.1769}}


\bibitem[Duval and Wicklund(1972)]%
        {duval1972theory}
\bibfield{author}{\bibinfo{person}{T.~Shelley Duval} {and} \bibinfo{person}{Robert~A. Wicklund}.} \bibinfo{year}{1972}\natexlab{}.
\newblock \bibinfo{booktitle}{\emph{A Theory of Objective Self-Awareness}}.
\newblock \bibinfo{publisher}{Academic Press}, \bibinfo{address}{New York, NY, USA}.
\newblock


\bibitem[Federoff and Harvey(1976)]%
        {FEDEROFF1976336}
\bibfield{author}{\bibinfo{person}{Nancy~A Federoff} {and} \bibinfo{person}{John~H Harvey}.} \bibinfo{year}{1976}\natexlab{}.
\newblock \showarticletitle{Focus of attention, self-esteem, and the attribution of causality}.
\newblock \bibinfo{journal}{\emph{Journal of Research in Personality}} \bibinfo{volume}{10}, \bibinfo{number}{3} (\bibinfo{year}{1976}), \bibinfo{pages}{336--345}.
\newblock
\showISSN{0092-6566}
\href{https://doi.org/10.1016/0092-6566(76)90023-4}{doi:\nolinkurl{10.1016/0092-6566(76)90023-4}}


\bibitem[Govern and Marsch(2001)]%
        {GOVERN2001366}
\bibfield{author}{\bibinfo{person}{John~M. Govern} {and} \bibinfo{person}{Lisa~A. Marsch}.} \bibinfo{year}{2001}\natexlab{}.
\newblock \showarticletitle{Development and Validation of the Situational Self-Awareness Scale}.
\newblock \bibinfo{journal}{\emph{Consciousness and Cognition}} \bibinfo{volume}{10}, \bibinfo{number}{3} (\bibinfo{year}{2001}), \bibinfo{pages}{366--378}.
\newblock
\showISSN{1053-8100}
\href{https://doi.org/10.1006/ccog.2001.0506}{doi:\nolinkurl{10.1006/ccog.2001.0506}}


\bibitem[Grüning et~al\mbox{.}(2023)]%
        {doi:10.1073/pnas.2213114120}
\bibfield{author}{\bibinfo{person}{David~J. Grüning}, \bibinfo{person}{Frederik Riedel}, {and} \bibinfo{person}{Philipp Lorenz-Spreen}.} \bibinfo{year}{2023}\natexlab{}.
\newblock \showarticletitle{Directing smartphone use through the self-nudge app one sec}.
\newblock \bibinfo{journal}{\emph{Proceedings of the National Academy of Sciences}} \bibinfo{volume}{120}, \bibinfo{number}{8} (\bibinfo{year}{2023}), \bibinfo{pages}{e2213114120}.
\newblock
\href{https://doi.org/10.1073/pnas.2213114120}{doi:\nolinkurl{10.1073/pnas.2213114120}}


\bibitem[Haliburton et~al\mbox{.}(2024)]%
        {10.1145/3613904.3642370}
\bibfield{author}{\bibinfo{person}{Luke Haliburton}, \bibinfo{person}{David~Joachim Gr\"{u}ning}, \bibinfo{person}{Frederik Riedel}, \bibinfo{person}{Albrecht Schmidt}, {and} \bibinfo{person}{Na\dj{}a Terzimehi\'{c}}.} \bibinfo{year}{2024}\natexlab{}.
\newblock \showarticletitle{A Longitudinal In-the-Wild Investigation of Design Frictions to Prevent Smartphone Overuse}. In \bibinfo{booktitle}{\emph{Proceedings of the 2024 CHI Conference on Human Factors in Computing Systems}} (Honolulu, HI, USA) \emph{(\bibinfo{series}{CHI '24})}. \bibinfo{publisher}{Association for Computing Machinery}, \bibinfo{address}{New York, NY, USA}, Article \bibinfo{articleno}{243}, \bibinfo{numpages}{16}~pages.
\newblock
\showISBNx{9798400703300}
\href{https://doi.org/10.1145/3613904.3642370}{doi:\nolinkurl{10.1145/3613904.3642370}}


\bibitem[Hiniker et~al\mbox{.}(2016)]%
        {10.1145/2858036.2858403}
\bibfield{author}{\bibinfo{person}{Alexis Hiniker}, \bibinfo{person}{Sungsoo~(Ray) Hong}, \bibinfo{person}{Tadayoshi Kohno}, {and} \bibinfo{person}{Julie~A. Kientz}.} \bibinfo{year}{2016}\natexlab{}.
\newblock \showarticletitle{MyTime: Designing and Evaluating an Intervention for Smartphone Non-Use}. In \bibinfo{booktitle}{\emph{Proceedings of the 2016 CHI Conference on Human Factors in Computing Systems}} (San Jose, California, USA) \emph{(\bibinfo{series}{CHI '16})}. \bibinfo{publisher}{Association for Computing Machinery}, \bibinfo{address}{New York, NY, USA}, \bibinfo{pages}{4746–4757}.
\newblock
\showISBNx{9781450333627}
\href{https://doi.org/10.1145/2858036.2858403}{doi:\nolinkurl{10.1145/2858036.2858403}}


\bibitem[Ickes et~al\mbox{.}(1973)]%
        {ICKES1973202}
\bibfield{author}{\bibinfo{person}{William~John Ickes}, \bibinfo{person}{Robert~A. Wicklund}, {and} \bibinfo{person}{C.Brian Ferris}.} \bibinfo{year}{1973}\natexlab{}.
\newblock \showarticletitle{Objective self awareness and self esteem}.
\newblock \bibinfo{journal}{\emph{Journal of Experimental Social Psychology}} \bibinfo{volume}{9}, \bibinfo{number}{3} (\bibinfo{year}{1973}), \bibinfo{pages}{202--219}.
\newblock
\showISSN{0022-1031}
\href{https://doi.org/10.1016/0022-1031(73)90010-3}{doi:\nolinkurl{10.1016/0022-1031(73)90010-3}}


\bibitem[Kim et~al\mbox{.}(2019)]%
        {10.1145/3314403}
\bibfield{author}{\bibinfo{person}{Jaejeung Kim}, \bibinfo{person}{Hayoung Jung}, \bibinfo{person}{Minsam Ko}, {and} \bibinfo{person}{Uichin Lee}.} \bibinfo{year}{2019}\natexlab{}.
\newblock \showarticletitle{GoalKeeper: Exploring Interaction Lockout Mechanisms for Regulating Smartphone Use}.
\newblock \bibinfo{journal}{\emph{Proc. ACM Interact. Mob. Wearable Ubiquitous Technol.}} \bibinfo{volume}{3}, \bibinfo{number}{1}, Article \bibinfo{articleno}{16} (\bibinfo{date}{March} \bibinfo{year}{2019}), \bibinfo{numpages}{29}~pages.
\newblock
\href{https://doi.org/10.1145/3314403}{doi:\nolinkurl{10.1145/3314403}}


\bibitem[Kim and Song(2024)]%
        {kim2024voice}
\bibfield{author}{\bibinfo{person}{Jieun Kim} {and} \bibinfo{person}{Hayeon Song}.} \bibinfo{year}{2024}\natexlab{}.
\newblock \showarticletitle{My Voice as a Daily Reminder: Self-Voice Alarm for Daily Goal Achievement}. In \bibinfo{booktitle}{\emph{Proceedings of the 2024 CHI Conference on Human Factors in Computing Systems}} (Honolulu, HI, USA) \emph{(\bibinfo{series}{CHI '24})}. \bibinfo{publisher}{Association for Computing Machinery}, \bibinfo{address}{New York, NY, USA}, Article \bibinfo{articleno}{716}, \bibinfo{numpages}{16}~pages.
\newblock
\showISBNx{9798400703300}
\href{https://doi.org/10.1145/3613904.3641932}{doi:\nolinkurl{10.1145/3613904.3641932}}


\bibitem[Liao et~al\mbox{.}(2024)]%
        {LIAO2024}
\bibfield{author}{\bibinfo{person}{Junyun Liao}, \bibinfo{person}{Siying He}, \bibinfo{person}{Wenting Feng}, {and} \bibinfo{person}{Raffaele Filieri}.} \bibinfo{year}{2024}\natexlab{}.
\newblock \showarticletitle{“I Love It” Versus “I Recommend It”: The Impact of Implicit and Explicit Endorsement Styles on Electronic Word-of-Mouth Persuasiveness}.
\newblock \bibinfo{journal}{\emph{Journal of Travel Research}} \bibinfo{volume}{63}, \bibinfo{number}{4} (\bibinfo{year}{2024}), \bibinfo{pages}{779--795}.
\newblock
\href{https://doi.org/10.1177/00472875231175083}{doi:\nolinkurl{10.1177/00472875231175083}}


\bibitem[Lyngs et~al\mbox{.}(2020)]%
        {10.1145/3313831.3376672}
\bibfield{author}{\bibinfo{person}{Ulrik Lyngs}, \bibinfo{person}{Kai Lukoff}, \bibinfo{person}{Petr Slovak}, \bibinfo{person}{William Seymour}, \bibinfo{person}{Helena Webb}, \bibinfo{person}{Marina Jirotka}, \bibinfo{person}{Jun Zhao}, \bibinfo{person}{Max Van~Kleek}, {and} \bibinfo{person}{Nigel Shadbolt}.} \bibinfo{year}{2020}\natexlab{}.
\newblock \showarticletitle{'I Just Want to Hack Myself to Not Get Distracted': Evaluating Design Interventions for Self-Control on Facebook}. In \bibinfo{booktitle}{\emph{Proceedings of the 2020 CHI Conference on Human Factors in Computing Systems}} (Honolulu, HI, USA) \emph{(\bibinfo{series}{CHI '20})}. \bibinfo{publisher}{Association for Computing Machinery}, \bibinfo{address}{New York, NY, USA}, \bibinfo{pages}{1–15}.
\newblock
\showISBNx{9781450367080}
\href{https://doi.org/10.1145/3313831.3376672}{doi:\nolinkurl{10.1145/3313831.3376672}}


\bibitem[Morita et~al\mbox{.}(2013)]%
        {morita2014anterior}
\bibfield{author}{\bibinfo{person}{Tomoyo Morita}, \bibinfo{person}{Hiroki~C. Tanabe}, \bibinfo{person}{Akihiro~T. Sasaki}, \bibinfo{person}{Koji Shimada}, \bibinfo{person}{Ryusuke Kakigi}, {and} \bibinfo{person}{Norihiro Sadato}.} \bibinfo{year}{2013}\natexlab{}.
\newblock \showarticletitle{The anterior insular and anterior cingulate cortices in emotional processing for self-face recognition}.
\newblock \bibinfo{journal}{\emph{Social Cognitive and Affective Neuroscience}} \bibinfo{volume}{9}, \bibinfo{number}{5} (\bibinfo{date}{03} \bibinfo{year}{2013}), \bibinfo{pages}{570--579}.
\newblock
\showISSN{1749-5016}
\href{https://doi.org/10.1093/scan/nst011}{doi:\nolinkurl{10.1093/scan/nst011}}


\bibitem[Murray et~al\mbox{.}(2015)]%
        {https://doi.org/10.1002/hbm.22703}
\bibfield{author}{\bibinfo{person}{Ryan~J. Murray}, \bibinfo{person}{Martin Debbané}, \bibinfo{person}{Peter~T. Fox}, \bibinfo{person}{Danilo Bzdok}, {and} \bibinfo{person}{Simon~B. Eickhoff}.} \bibinfo{year}{2015}\natexlab{}.
\newblock \showarticletitle{Functional connectivity mapping of regions associated with self- and other-processing}.
\newblock \bibinfo{journal}{\emph{Human Brain Mapping}} \bibinfo{volume}{36}, \bibinfo{number}{4} (\bibinfo{year}{2015}), \bibinfo{pages}{1304--1324}.
\newblock
\href{https://doi.org/10.1002/hbm.22703}{doi:\nolinkurl{10.1002/hbm.22703}}


\bibitem[{Oxford University Press}(2024)]%
        {oxford2024brainrot}
\bibfield{author}{\bibinfo{person}{{Oxford University Press}}.} \bibinfo{year}{2024}\natexlab{}.
\newblock \bibinfo{booktitle}{\emph{'Brain rot' named Oxford Word of the Year 2024}}.
\newblock
\urldef\tempurl%
\url{https://corp.oup.com/news/brain-rot-named-oxford-word-of-the-year-2024/}
\showURL{%
Retrieved September 10, 2025 from \tempurl}


\bibitem[Park(2024)]%
        {PARK2024102594}
\bibfield{author}{\bibinfo{person}{Eunil Park}.} \bibinfo{year}{2024}\natexlab{}.
\newblock \showarticletitle{Examining metaverse game platform adoption: Insights from innovation, behavior, and coolness}.
\newblock \bibinfo{journal}{\emph{Technology in Society}}  \bibinfo{volume}{77} (\bibinfo{year}{2024}), \bibinfo{pages}{102594}.
\newblock
\showISSN{0160-791X}
\href{https://doi.org/10.1016/j.techsoc.2024.102594}{doi:\nolinkurl{10.1016/j.techsoc.2024.102594}}


\bibitem[Picco et~al\mbox{.}(2025)]%
        {PICCO2025421}
\bibfield{author}{\bibinfo{person}{Angèle Picco}, \bibinfo{person}{Arjan Stuiver}, \bibinfo{person}{Joost {De Winter}}, {and} \bibinfo{person}{Dick {De Waard}}.} \bibinfo{year}{2025}\natexlab{}.
\newblock \showarticletitle{“Why were you speeding?”: A self-confrontation study on awareness and reasons for speed behaviour}.
\newblock \bibinfo{journal}{\emph{Transportation Research Part F: Traffic Psychology and Behaviour}}  \bibinfo{volume}{109} (\bibinfo{year}{2025}), \bibinfo{pages}{421--438}.
\newblock
\showISSN{1369-8478}
\href{https://doi.org/10.1016/j.trf.2024.12.015}{doi:\nolinkurl{10.1016/j.trf.2024.12.015}}


\bibitem[Qin and Northoff(2011)]%
        {QIN20111221}
\bibfield{author}{\bibinfo{person}{Pengmin Qin} {and} \bibinfo{person}{Georg Northoff}.} \bibinfo{year}{2011}\natexlab{}.
\newblock \showarticletitle{How is our self related to midline regions and the default-mode network?}
\newblock \bibinfo{journal}{\emph{NeuroImage}} \bibinfo{volume}{57}, \bibinfo{number}{3} (\bibinfo{year}{2011}), \bibinfo{pages}{1221--1233}.
\newblock
\showISSN{1053-8119}
\href{https://doi.org/10.1016/j.neuroimage.2011.05.028}{doi:\nolinkurl{10.1016/j.neuroimage.2011.05.028}}
\newblock
\shownote{Special Issue: Educational Neuroscience}.


\bibitem[Schwarzer and Renner(2009)]%
        {Ralf2009}
\bibfield{author}{\bibinfo{person}{Ralf Schwarzer} {and} \bibinfo{person}{Britta Renner}.} \bibinfo{year}{2009}\natexlab{}.
\newblock \showarticletitle{Health-Specific Self-Efficacy Scales}.
\newblock  (\bibinfo{date}{01} \bibinfo{year}{2009}).
\newblock


\bibitem[Silvia and Duval(2001)]%
        {Silvia2001Objective}
\bibfield{author}{\bibinfo{person}{Paul~J. Silvia} {and} \bibinfo{person}{T.~Shelley Duval}.} \bibinfo{year}{2001}\natexlab{}.
\newblock \showarticletitle{Objective Self-Awareness Theory: Recent Progress and Enduring Problems}.
\newblock \bibinfo{journal}{\emph{Personality and Social Psychology Review}} \bibinfo{volume}{5}, \bibinfo{number}{3} (\bibinfo{year}{2001}), \bibinfo{pages}{230--241}.
\newblock
\href{https://doi.org/10.1207/S15327957PSPR0503\_4}{doi:\nolinkurl{10.1207/S15327957PSPR0503\_4}}


\bibitem[Venkatesh and Davis(2000)]%
        {doi:10.1287/mnsc.46.2.186.11926}
\bibfield{author}{\bibinfo{person}{Viswanath Venkatesh} {and} \bibinfo{person}{Fred~D. Davis}.} \bibinfo{year}{2000}\natexlab{}.
\newblock \showarticletitle{A Theoretical Extension of the Technology Acceptance Model: Four Longitudinal Field Studies}.
\newblock \bibinfo{journal}{\emph{Management Science}} \bibinfo{volume}{46}, \bibinfo{number}{2} (\bibinfo{year}{2000}), \bibinfo{pages}{186--204}.
\newblock
\href{https://doi.org/10.1287/mnsc.46.2.186.11926}{doi:\nolinkurl{10.1287/mnsc.46.2.186.11926}}


\bibitem[WITTE(1996)]%
        {WITTE01101996}
\bibfield{author}{\bibinfo{person}{KIM WITTE}.} \bibinfo{year}{1996}\natexlab{}.
\newblock \showarticletitle{Predicting Risk Behaviors: Development and Validation of a Diagnostic Scale}.
\newblock \bibinfo{journal}{\emph{Journal of Health Communication}} \bibinfo{volume}{1}, \bibinfo{number}{4} (\bibinfo{year}{1996}), \bibinfo{pages}{317--342}.
\newblock
\href{https://doi.org/10.1080/108107396127988}{doi:\nolinkurl{10.1080/108107396127988}}


\bibitem[Wixom and Todd(2005)]%
        {662ada62-a334-3561-89a0-428c9f338300}
\bibfield{author}{\bibinfo{person}{Barbara~H. Wixom} {and} \bibinfo{person}{Peter~A. Todd}.} \bibinfo{year}{2005}\natexlab{}.
\newblock \showarticletitle{A Theoretical Integration of User Satisfaction and Technology Acceptance}.
\newblock \bibinfo{journal}{\emph{Information Systems Research}} \bibinfo{volume}{16}, \bibinfo{number}{1} (\bibinfo{year}{2005}), \bibinfo{pages}{85--102}.
\newblock
\showISSN{10477047, 15265536}
\urldef\tempurl%
\url{http://www.jstor.org/stable/23015766}
\showURL{%
\tempurl}


\bibitem[Wu et~al\mbox{.}(2024)]%
        {10.1145/3613904.3642790}
\bibfield{author}{\bibinfo{person}{Ruolan Wu}, \bibinfo{person}{Chun Yu}, \bibinfo{person}{Xiaole Pan}, \bibinfo{person}{Yujia Liu}, \bibinfo{person}{Ningning Zhang}, \bibinfo{person}{Yue Fu}, \bibinfo{person}{Yuhan Wang}, \bibinfo{person}{Zhi Zheng}, \bibinfo{person}{Li Chen}, \bibinfo{person}{Qiaolei Jiang}, \bibinfo{person}{Xuhai Xu}, {and} \bibinfo{person}{Yuanchun Shi}.} \bibinfo{year}{2024}\natexlab{}.
\newblock \showarticletitle{MindShift: Leveraging Large Language Models for Mental-States-Based Problematic Smartphone Use Intervention}. In \bibinfo{booktitle}{\emph{Proceedings of the 2024 CHI Conference on Human Factors in Computing Systems}} (Honolulu, HI, USA) \emph{(\bibinfo{series}{CHI '24})}. \bibinfo{publisher}{Association for Computing Machinery}, \bibinfo{address}{New York, NY, USA}, Article \bibinfo{articleno}{248}, \bibinfo{numpages}{24}~pages.
\newblock
\showISBNx{9798400703300}
\href{https://doi.org/10.1145/3613904.3642790}{doi:\nolinkurl{10.1145/3613904.3642790}}


\bibitem[{YouTube Official Blog}(2025)]%
        {youtube2025cannes}
\bibfield{author}{\bibinfo{person}{{YouTube Official Blog}}.} \bibinfo{year}{2025}\natexlab{}.
\newblock \bibinfo{booktitle}{\emph{Neal Mohan at Cannes Lions 2025: What 20 years of YouTube reveals about creativity’s future}}.
\newblock
\urldef\tempurl%
\url{https://blog.youtube/news-and-events/neal-mohan-cannes-2025/}
\showURL{%
Retrieved Accessed: 2025-09-10 from \tempurl}


\bibitem[Zhang et~al\mbox{.}(2019)]%
        {ZHANG2019101243}
\bibfield{author}{\bibinfo{person}{Xing Zhang}, \bibinfo{person}{You Wu}, {and} \bibinfo{person}{Shan Liu}.} \bibinfo{year}{2019}\natexlab{}.
\newblock \showarticletitle{Exploring short-form video application addiction: Socio-technical and attachment perspectives}.
\newblock \bibinfo{journal}{\emph{Telematics and Informatics}}  \bibinfo{volume}{42} (\bibinfo{year}{2019}), \bibinfo{pages}{101243}.
\newblock
\showISSN{0736-5853}
\href{https://doi.org/10.1016/j.tele.2019.101243}{doi:\nolinkurl{10.1016/j.tele.2019.101243}}


\end{thebibliography}
